\documentclass[aps,pra,10pt,reprint,superscriptaddress]{revtex4-2}
\usepackage{amsmath,amsfonts,amssymb}
\usepackage{xcolor,graphicx}
\usepackage{mwe}
\usepackage{overpic}
\usepackage{braket}
\usepackage[pdftex]{hyperref}
\hypersetup{
    colorlinks = true,
    linkcolor = cyan,
    anchorcolor =cyan,
    citecolor = cyan,
    filecolor = cyan,
    urlcolor = cyan}

\begin{document}
\title{From Rings to Top-Hat beams}

\author{M. A. Jácome-Silva}
\affiliation{Instituto Nacional de Astrofísica Óptica y Electrónica\\ Luis Enrique Erro 1, Santa María Tonantzintla, Puebla, 72840, Mexico}

\author{I. Julían-Macías}
\email{isjuma@inaoep.mx}
\affiliation{Instituto Nacional de Astrofísica Óptica y Electrónica\\ Luis Enrique Erro 1, Santa María Tonantzintla, Puebla, 72840, Mexico}

\author{I. Ramos-Prieto}
\email{iran@inaoep.mx}
\affiliation{Instituto Nacional de Astrofísica Óptica y Electrónica\\ Luis Enrique Erro 1, Santa María Tonantzintla, Puebla, 72840, Mexico}

\author{U. Ruíz-Corona}
\affiliation{Instituto Nacional de Astrofísica Óptica y Electrónica\\ Luis Enrique Erro 1, Santa María Tonantzintla, Puebla, 72840, Mexico}

\author{F. Soto-Eguibar}
\affiliation{Instituto Nacional de Astrofísica Óptica y Electrónica\\ Luis Enrique Erro 1, Santa María Tonantzintla, Puebla, 72840, Mexico}

\author{D. {Sánchez}-{de-la-Llave}}
\affiliation{Instituto Nacional de Astrofísica Óptica y Electrónica\\ Luis Enrique Erro 1, Santa María Tonantzintla, Puebla, 72840, Mexico}

\author{H. M. Moya-Cessa}
\affiliation{Instituto Nacional de Astrofísica Óptica y Electrónica\\ Luis Enrique Erro 1, Santa María Tonantzintla, Puebla, 72840, Mexico}

\begin{abstract}
We present the exact analytical paraxial propagation of structured light beams that transition from Ring annular profiles to top-hat intensity distributions. The initial field is defined as a superposition of a Gaussian-weighted power-law core and a singular inverse-quadratic modulation term, both carrying an azimuthal phase factor. By solving the Fresnel diffraction integral in cylindrical coordinates, we obtain exact closed-form expressions for the propagated field at arbitrary planes. The paraxial evolution is shown to be governed by a Cauchy-Riemann beam term and an infinite series of modified Bessel functions of the second kind weighted by an azimuthal phase factor. This analytical framework demonstrates how tuning the source parameters enables a continuous transition from ring-dominated annular profiles to uniform top-hat beams. For the fundamental mode ($l=0$), the singular component fills the central intensity null, producing a flat transverse plateau.
\end{abstract}

\maketitle
Light beams with a helical wavefront $\exp(il\theta)$ carry orbital angular momentum proportional to the topological charge $l$ \cite{Allen_1992}. These fields are characterized by an on-axis phase singularity yielding a central intensity null, a feature that is advantageous for applications such as optical particle manipulation and high-capacity information encoding~\cite{Ashkin_1997,Padgett_2011}. However, many applications in material processing, laser surgery, and quantum technologies require the opposite: uniform transverse intensity profiles, commonly referred to as flat-top or top-hat beams~\cite{Gori_1994}. Unlike Gaussian beams, which can cause thermal damage due to their sharp central intensity peak, a top-hat beam delivers constant irradiance across its illuminated area. This distribution is used to ensure consistent depth in laser ablation \cite{Russo_2002}, controlled thermal deposition in laser annealing~\cite{Wood_1981}. In quantum science, top-hat beams are used to generate uniform optical traps (box potentials) for studying Bose-Einstein condensates \cite{Gaunt_2013} and for creating ordered atomic arrays \cite{Barredo_2016,Mielec_2018}.

Analytical modeling of top-hat beam propagation is typically performed by approximating these profiles with high-order super-Gaussian functions or series expansions of flattened Gaussian beams \cite{Gori_1994, Palma_1996, Bagini_1996,Ibnchaikh_2001}. However, the paraxial propagation of super-Gaussian beams cannot be represented in terms of elementary functions, requiring either split-step Fourier numerical simulations or modal expansions with a high number of terms \cite{Palma_1996}. These approaches are computationally intensive and prone to spatial frequency aliasing. We remark that in previous works \cite{Gori_1994, Palma_1996, Bagini_1996,Ibnchaikh_2001, Palma_1996}, top-hat profiles were generated at $z = 0$, and no analytical or numerical results were presented for the generation of top-hat profiles at $z>0$. While experimental conversion of Gaussian beams into top-hat distributions can be achieved using diffractive optical elements, spatial light modulators, or dielectric metalenses \cite{Chen_2025, Pal_2018}, the lack of closed-form propagation formulas limits the systematic design and real-time optimization of these optical systems.

To address these limitations, we develop an exact analytical paraxial propagation framework for a family of structured light beams that continuously interpolate between ring-shaped and top-hat intensity distributions for arbitrary topological charges. The initial field is defined as the superposition of a Gaussian-weighted power-law core modulated by an azimuthal phase factor---which gives rise to a Cauchy-Riemann beam upon propagation~\cite{CRB_2024,CRB_GRIN_2024,ACRB_2025}---and a term with a singular inverse-quadratic phase factor carrying an azimuthal phase factor---yielding a series representation in terms of modified Bessel functions of the second kind, multiplied by an azimuthal phase factor---. By evaluating the Fresnel diffraction integral in cylindrical coordinates, exact closed-form solutions are obtained for both components. The total propagated field is expressed as the sum of a Cauchy-Riemann term and an infinite series of modified Bessel functions of the second kind modulated by an azimuthal phase factor. Tuning the source parameters allows continuous transition between ring-shaped profiles and flat plateaus (top-hat beams); in the fundamental mode ($l=0$), the singular term fills the central intensity null, producing a uniform top-hat profile.

In the paraxial regime, the propagation of a monochromatic scalar electric field $E(x,y;z)$ through free space along the optical axis $z$ is governed by the paraxial Helmholtz equation. Within this framework, the field at any propagation distance $z$ is obtained by applying the paraxial propagation operator to the initial field distribution $E(x,y;0)$ at the source plane $z=0$~\cite{Stoler_1981}:
\begin{equation}\label{E_z}
    E(x,y;z) = e^{\tfrac{iz}{2k}\nabla_{\perp}^2}E(x,y;0),
\end{equation}
where $\nabla_{\perp}^2 = \partial^2/\partial x^2 + \partial^2/\partial y^2$ is the transverse Laplacian operator, $k = 2\pi/\lambda$ is the wavenumber, and $\lambda$ is the wavelength. For our purposes, we consider the initial field distribution
\begin{equation}\label{E_0}
    E(r, \theta; 0) = e^{-ar^2}\left(\alpha\,r^l + \beta e^{-\tfrac{b}{r^2}}\right)e^{il\theta},
\end{equation}
where $a$, $b$, $\alpha$, $\beta$ are real constants, $(r,\theta)$ are polar coordinates, and $l$ is a positive integer representing the topological charge that encodes the orbital angular momentum of the field. The first term of the initial condition in Eq.~\eqref{E_0}, when acted upon by the paraxial propagation operator in Eq.~\eqref{E_z}, admits an exact solution via an operator algebra approach and is closely related to the family of Cauchy-Riemann beams \cite{CRB_2024}. This nomenclature stems from the fact that the transverse modulation factor $(x+iy)^l = r^l \exp(il\theta)$ satisfies the Cauchy-Riemann equations, representing an entire function of the complex variable. By leveraging the algebraic properties of the paraxial propagation operator, the exact evolution of this Gaussian-modulated entire function at a propagation distance $z$ is given by~\cite{CRB_2024}:
\begin{equation}\label{E_alpha}
\begin{split}
    e^{\tfrac{iz}{2k}\nabla_{\perp}^2}\left(e^{-ar^2}r^l e^{il\theta}\right)&= \frac{1}{w(z)}e^{-\tfrac{ar^2}{w(z)}}\left(\frac{x+iy}{w(z)}\right)^l\\
    &=\frac{1}{w(z)}e^{-\tfrac{ar^2}{w(z)}}\left(\frac{r}{w(z)}\right)^le^{il\theta},
\end{split}
\end{equation}
with the complex scaling parameter $w(z) = 1 + i2az/k$.

While the operator-based method provides an elegant and direct solution for the core entire-function component, the paraxial propagation of the complete initial field in Eq.~\eqref{E_0}---which includes the singular, non-entire inverse-quadratic term---is more naturally handled via the Fresnel diffraction integral, since the $1/r^2$ term does not close a Lie algebra under commutation with $\nabla_\perp^2$. In this representation, paraxial propagation is described by the classical two-dimensional Fresnel diffraction integral. Applying the Anger-Jacobi identity~\cite{GradShteyn}, $\exp\left(i x \cos \psi\right) = \sum i^m J_m(x) \exp\left(im\psi\right)$, together with $J_m(-x) = (-1)^m J_m(x)$, and integrating over the azimuthal coordinate yields
\begin{equation}
\begin{split}
    e^{\tfrac{iz}{2k}\nabla_{\perp}^2}\left(e^{-ar^2}e^{-\tfrac{b}{r^2}}e^{il\theta}\right) &= \frac{k}{z}(-i)^{l+1} e^{\tfrac{ik r^2}{2z}} e^{il\theta}\\
    &\times \int\limits_{0}^{\infty} e^{-\gamma\rho^2}e^{-\tfrac{b}{\rho^2}} J_l\left(\frac{kr\rho}{z}\right)  \rho d\rho,
\end{split}
\end{equation}
with $\gamma = a - ik/(2z)$. To evaluate the remaining radial integral, we express the Bessel function in terms of its power series representation, allowing the integration over the radial variable to be written as:
\begin{equation}\label{integral_rho}
\begin{split}
  &\int\limits_{0}^{\infty} e^{-\gamma\rho^2}e^{-\tfrac{b}{\rho^2}} J_l\left(\frac{kr\rho}{z}\right)  \rho d\rho\\
  & = \sum\limits_{m=0}^{\infty} \frac{(-1)^m}{m! (m+l)!} \left(\frac{k r}{2z}\right)^{2m+l}\int\limits_{0}^{\infty} e^{-\gamma\rho^2}e^{-\tfrac{b}{\rho^2}} \rho^{2m+l}  \rho d\rho.
\end{split}
\end{equation}
By performing the change of variable $u = \rho^2$, we can evaluate this integral by leveraging the following identity from Gradshteyn and Ryzhik~\cite{GradShteyn}:
\begin{equation}\label{relacion_Gradshteyn}
\begin{split}
    \frac{1}{2}\int\limits_{0}^{\infty} e^{-\gamma u}e^{-\tfrac{b}{u}}& u^{m + \frac{l}{2}} du 
    \\&= \left(\frac{b}{\gamma}\right)^{\tfrac{m}{2} + \tfrac{l}{4} + \tfrac{1}{2}} K_{\tfrac{m}{2} + \tfrac{l}{2} + 1}\left[2\sqrt{b \gamma}\right],
\end{split}
\end{equation}
where $K_\nu(x)$ is the modified Bessel function of the second kind of order $\nu$. Finally, by combining these results and expressing $\gamma$ in terms of the scaling parameter $w(z)$, the propagated field is exactly given by:
\begin{equation}\label{E_beta}
\begin{split}
   &e^{\tfrac{iz}{2k}\nabla_{\perp}^2}\left(e^{-ar^2}e^{-\tfrac{b}{r^2}}e^{il\theta}\right) = \frac{k}{z}(-i)^{l+1} e^{\tfrac{ik r^2}{2z}} e^{il\theta}\\&\times\sum\limits_{m=0}^{\infty} \frac{(-1)^m}{m! (m+l)!} \left(\frac{k r}{2z}\right)^{2m+l} \left(\frac{2ibz}{k w(z)}\right)^{\tfrac{m}{2} + \tfrac{l}{4} + \tfrac{1}{2}}\\&\times K_{\tfrac{m}{2} + \tfrac{l}{2} +1}\left[2\sqrt{-\frac{ibk w(z)}{2z}}\right].
\end{split}
\end{equation}
Consequently, exploiting the linearity of the paraxial propagation operator and superposing the solutions (Eqs.~\eqref{E_alpha} and \eqref{E_beta}), the total propagated field $E(r,\theta;z)$ associated with the initial condition Eq.~\eqref{E_0} reads:
\begin{equation}\label{E_total}
    E(r,\theta;z) = \alpha E_{\alpha}(r,\theta;z) - \beta E_{\beta}(r,\theta;z),
\end{equation}
where $E_{\alpha}$ and $E_{\beta}$ denote, respectively, the paraxial evolution of the Gaussian-weighted power-law core and of the singular inverse-quadratic modulation component, both carrying an azimuthal phase factor. Physically, the total field $E(r,\theta;z)$ in Eq.~\eqref{E_total} is a superposition of two distinct beam families. The term $E_{\alpha}$, Eq.~\eqref{E_alpha}, corresponds to a Cauchy-Riemann beam that retains its vortex structure while scaling self-similarly with the complex beam parameter $w(z)$. The term $E_{\beta}$, Eq.~\eqref{E_beta}, encodes the contribution of the singular inverse-quadratic modulation, manifesting as a series of modified Bessel functions that concentrate energy toward the optical axis. This series converges rapidly due to the factorial growth of the terms $m!(m+l)!$ in the denominator of Eq.~\eqref{E_beta}, meaning that only a few terms are required for accurate numerical computation. Without loss of generality, we adopt the negative sign in Eq.~\eqref{E_total}; the positive sign simply inverts the relative phase and yields equivalent dynamics. The relative amplitudes $\alpha$ and $\beta$ control the interference between these components, enabling continuous tuning from a hollow vortex profile to a flat-top (top-hat) intensity distribution for the fundamental mode ($l=0$). This mechanism underlies the formation of Ring beams with adjustable central filling and orbital angular momentum.

To validate the analytical expressions obtained, we experimentally generated these structured fields using a spatial light modulator. The setup employs a standard $4f$ optical system illuminated by a collimated He-Ne laser ($\lambda = 632.8$ nm), where the amplitude and phase profiles of the initial condition, Eq.~\eqref{E_0}, are encoded via a synthetic phase-only hologram ~\cite{Arrizon_2007}. By spatially filtering the beam at the Fourier plane of the first lens to isolate the first diffraction order, the desired field is reconstructed and its propagated intensity distributions are recorded at various planes along the optical axis using a CCD camera. In what follows, we analyze several representative cases of this family of structured beams.\\

\begin{figure}[h!]
\centering\includegraphics[width=\linewidth]{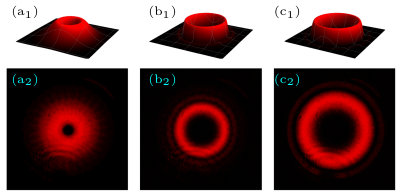}
\caption{Transverse intensity distribution $|E(r,\theta; z)|^2$ of the total field in Eq.~\eqref{E_total} for a topological charge $l = 5$ at propagation distances $z = 0.0$, $0.5$, and $1.0~\text{m}$. The panels ($\mathrm{a}_1$)-($\mathrm{c}_1$) show the intensity in 3D surface plots obtained analytically, while the panels ($\mathrm{a}_2$)-($\mathrm{c}_2$) show the 2D transverse intensity obtained experimentally. The experimental parameters are: $a = 0.1 \times 10^7 \, \mathrm{m}^{-2}$, $b = 0.5 \times 10^7 \, \mathrm{m}^2$, $\alpha = 1.0~\text{V/m}$, $\beta = 1.0~\text{V/m}$, $r_0 = 1~\text{m}$ (where $r_0$ is a scale parameter introduced in the numerical implementation of the hologram to maintain dimensional consistency of the physical coordinates). The observation window is $4~\text{mm} \times 4~\text{mm} $.}
\label{figura1}
\end{figure}

\textit{Ring beams with topological charge.} We first consider $l \neq 0$, where the initial field carries orbital angular momentum. Figure~\ref{figura1} shows the transverse intensity of the field (Eq.~\eqref{E_total}) for charge $l=5$ at $z=0.0$, $0.5$, and $1.0~\text{m}$. Panels ($\mathrm{a}_1$)-($\mathrm{c}_1$) present the $3D$ intensity distribution $|E(r,\theta;z)|^2$, while ($\mathrm{a}_2$)-($\mathrm{c}_2$) display the corresponding experimental patterns. The intensity distribution results from the interference between the Cauchy-Riemann term $E_{\alpha}$, which carries the vortex structure and scales by $w(z)$, and the Bessel term $E_{\beta}$, which concentrates energy axially via the modified Bessel functions $K_\nu$. For $l>0$, the phase singularity at $r=0$ forces the field amplitude to vanish, producing a pronounced annular profile with a dark core. As the beam propagates, diffraction increases the ring radius while the peak intensity diminishes as energy spreads radially. The vortex topology remains topologically protected throughout propagation, as the orbital angular momentum is conserved~\cite{JacomeOperator}. However, the Bessel contribution $E_{\beta}$ partially fills the central region at intermediate distances, modulating the annular contrast without destroying the dark core. The analytical model shows an excellent qualitative agreement with the experimental results, reproducing all key spatial features, including the central dark core size and the radial expansion during propagation.

\begin{figure}[h!]
\centering\includegraphics[width=\linewidth]{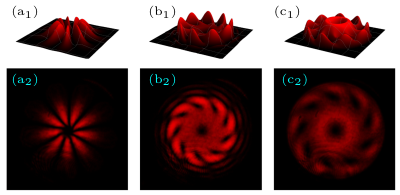}
\caption{Transverse intensity distribution $|E_{+}(r,\theta;z)|^2$ for the superposition of two Ring beams with topological charges $l_1=1$ and $l_2=10$ at $z=0.0$, $0.5$, and $1.0~\text{m}$. Experimental parameters same as Fig.~\ref{figura1}.}
\label{figura2}
\end{figure}

\textit{Superposition of two topological charges.} We next examine the propagation of a linear superposition of two Ring beams carrying distinct topological charges $l_1=1$ and $l_2=10$:
\begin{equation}\label{E_super}
    E_{\pm}(r,\theta;z) = E(r,\theta;z;l_1) \pm E(r,\theta;z;l_2),
\end{equation}
using the same parameters as in Fig.~\ref{figura1}. Figure~\ref{figura2} presents the propagation of the superposition $E_{+} = E(l_1) + E(l_2)$. At the source plane $z=0$, the azimuthal interference between the helical phases produces a petal-like pattern with $|l_1 - l_2| = 9$ lobes. For $E_{+}$, the lobes align with the maxima of $\cos[(l_1-l_2)\theta]$, whereas for $E_{-}$ the pattern is rotated by $\pi/9$ rad. Under propagation, both components evolve independently according to Eq.~\eqref{E_total}: the lobes expand radially and broaden diffractively, yet the $9$-fold rotational symmetry remains strictly preserved because the relative topological charge $\Delta l = l_1-l_2$ is a conserved quantity. The analytical structure of Eq.~\eqref{E_total} enables evaluating the spatial redistribution of the orbital angular momentum density within the lobes directly, bypassing numerical Fresnel diffraction simulations.

\begin{figure}[h!]
\centering\includegraphics[width=\linewidth]{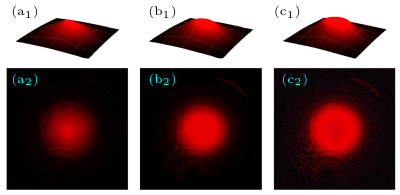}
\caption{Transverse intensity distribution $|E(r,\theta; z)|^2$ of the fundamental mode $l=0$ at propagation distances $z=0.0$, $0.5$, and $1.0~\text{m}$. The experimental parameters are: $a = 0.1\times 10^{6} \, \mathrm{m}^{-2}$, $b = 0.5 \times 10^{-6} \, \mathrm{m}^{2}$, $\alpha = 1.0~\text{V/m}$, and $\beta = 1.0~\text{V/m}$.}
\label{figura3}
\end{figure}

\begin{figure}[h!]
\centering\includegraphics[width=\linewidth]{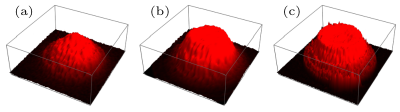}
\caption{Three-dimensional surface reconstructions of the experimentally recorded 2D transverse intensity distributions $|E(r,\theta;z)|^2$ of the fundamental mode ($l=0$) at propagation distances (a)~$z=0$, (b)~$z=0.5~\text{m}$, and (c)~$z=1.0~\text{m}$. Each surface is rendered directly from the corresponding CCD-captured intensity map (Fig.~\ref{figura3}, panels~$\mathrm{a}_2$--$\mathrm{c}_2$) by mapping the pixel grey-levels onto the vertical axis, providing a volumetric view of the transverse energy redistribution as the beam propagates. The progressive filling of the central region and the formation of a flat intensity plateau at $z=1.0~\text{m}$ are clearly resolved in 3D, confirming the transition to a top-hat profile predicted analytically by Eq.~\eqref{E_total}.}
\label{figura4}
\end{figure}

\textit{Top-hat regime: fundamental mode $l=0$.} Setting $l=0$ suppresses the helical phase singularity, allowing the singular term $E_{\beta}$ to fill the axial intensity null. Figure~\ref{figura3} shows the transverse intensity distribution $|E(r,\theta;z)|^2$ for the fundamental mode at propagation distances $z=0.0$, $0.5$, and $1.0~\text{m}$. Under this regime, the initial field in Eq.~\eqref{E_0} simplifies to $E(r,\theta;0) = \exp(-ar^2)[\alpha + \beta \exp(-b/r^2)]$, where the constant $\alpha$ provides a Gaussian pedestal and the term $\exp(-b/r^2)$ injects energy near the optical axis. The transverse profile is governed by the ratio $\beta/\alpha$: a low ratio preserves a ring-like dip, a moderate ratio flattens the central region into a uniform top-hat plateau, and a high ratio produces a centrally peaked beam. The optimal flat-top distance is physically determined by the balance between the Gaussian decay width $1/\sqrt{a}$ and the singular modulation range $\sqrt{b}$. During propagation, these two opposing effects compensate each other at a specific plane to form a flat, ripple-free intensity plateau that diffracts without developing the spatial oscillations typical of super-Gaussian approximations~\cite{Palma_1996}. Experimental measurements confirm this physical behavior, showing that the reconstructed spatial profiles match the analytical predictions. As shown in the three-dimensional reconstructions of Fig.~\ref{figura4}, the beam starts as a centrally peaked profile at $z=0$ (panel a), redistributes energy toward the axis at $z=0.5~\text{m}$ (panel b), and finally establishes the flat top-hat plateau at $z=1.0~\text{m}$ (panel c).

\textit{Conclusion.} We have derived and experimentally validated an exact closed-form analytical expression for the paraxial propagation of a family of structured light beams that continuously interpolates between ring-shaped vortex profiles and uniform top-hat distributions. The central result is the total propagated field $E(r,\theta;z)$ in Eq.~\eqref{E_total}, expressed as the superposition of a Cauchy-Riemann beam $E_{\alpha}$ (Eq.~\eqref{E_alpha}) that self-similarly scales with the complex parameter $w(z)$, and a singular modulation component $E_{\beta}$ (Eq.~\eqref{E_beta}) given by a rapidly convergent series of modified Bessel functions of the second kind. This decomposition is exact for any topological charge $l$ and any propagation distance $z$, requiring only the evaluation of elementary functions and a handful of Bessel series terms---a computational cost far below that of split-step Fourier simulations or super-Gaussian modal expansions.

The top-hat regime ($l=0$) deserves special emphasis. When the topological charge is suppressed, the factor $r^l$ in Eq.~\eqref{E_0} reduces to unity, and the singular term $\exp(-b/r^2)$ fills the central intensity null that characterizes vortex beams. Rather than starting with a top-hat profile at the source plane, the beam acquires a top-hat-like distribution at a distance controlled by the source parameters $a$ and $b$, after which the normal Gaussian beam eventually predominates. Figures~\ref{figura3} and \ref{figura4} show this evolution: at $z=0$ the profile is modulated, at an intermediate plane it becomes nearly flat, and beyond that it diffracts while preserving its uniformity. The plateau, once formed, propagates without developing the spatial oscillations that plague super-Gaussian approximations~\cite{Palma_1996}, and its width and edge steepness are independently controlled by $a$ and $b$, respectively. This offers the experimentalist two orthogonal knobs to place the top-hat at the desired working distance, bypassing iterative numerical optimization. This analytical control is particularly valuable for applications that demand uniform illumination with sharp edges, such as uniform optical dipole traps for cold atoms~\cite{Gaunt_2013, Barredo_2016}, where the beam uniformity directly determines the trapping potential flatness. Furthermore, beyond the $l=0$ case, the framework also captures the propagation of vortex ring beams (Fig.~\ref{figura1}) and their linear superpositions (Fig.~\ref{figura2}), demonstrating that the closed-form structure preserves topological features---dark cores, petal symmetries, and orbital angular momentum content---under free-space diffraction without numerical artifacts.

The model retains the algebraic simplicity of Gaussian-beam analysis while achieving the profile versatility of numerical beam-shaping methods, bridging a longstanding gap in structured-light theory. Because it is closed-form and grid-free, it is immediately compatible with real-time optimization loops and with inverse-design protocols for diffractive optics and metalens arrays~\cite{Chen_2025, Pal_2018}. Future extensions of this formalism could incorporate non-paraxial corrections for high-numerical-aperture focusing, account for atmospheric turbulence in long-range free-space links, or be adapted to partially coherent and polychromatic sources for applications in laser-material processing and biomedical imaging~\cite{Russo_2002}.

%
\end{document}